\documentclass[]{iopart}

\usepackage{iopams}
\usepackage{graphicx}
\usepackage{bm}

\newcommand{\s}{\sum\limits}
\newcommand{\p}{\prod\limits}

\newcommand{\il}{\int\limits}
\newcommand{\ba}{\begin{array}}
\newcommand{\ea}{\end{array}}
\newcommand{\lt}{\left}
\newcommand{\rt}{\right}

\newcommand{\ep}{\varepsilon}
\newcommand{\bb}{\boldsymbol}

\newcommand{\h}{^\dagger}

\newcommand{\im}{\mathrm{Im}\,}

\begin{document}

\title[Metallic proximity effect in ballistic graphene]
{Metallic proximity effect in ballistic graphene with resonant scatterers}

\author{M Titov$^1$  P M Ostrovsky$^{2,3}$ I V Gornyi$^{2,4}$}

\address{$^1$ School of Engineering \& Physical Sciences, 
Heriot-Watt University, Edinburgh EH14 4AS, UK}

\address{$^2$ Institut f\"ur Nanotechnologie, Forschungszentrum Karlsruhe,
76021 Karlsruhe, Germany}

\address{$^3$ L.~D.~Landau Institute for Theoretical Physics RAS,
119334 Moscow, Russia}

\address{$^4$ A.F.~Ioffe Physico-Technical Institute,
194021 St.~Petersburg, Russia}

\ead{m.titov@hw.ac.uk}

\begin{abstract}
We study the effect of resonant scatterers on the local density of states 
in a rectangular graphene setup with metallic leads. We find that 
the density of states in a vicinity of the Dirac point acquires 
a strong position dependence due to both metallic proximity effect
and impurity scattering. This effect may prevent uniform 
gating of weakly-doped samples. We also demonstrate that even 
a single-atom impurity may essentially alter electronic states 
at low-doping on distances of the order of the sample size 
from the impurity. 
\end{abstract}



\maketitle

\section{Introduction}

Since the discovery of graphene \cite{Novoselov04,Novoselov05} 
a great progress has been made in production of high quality samples 
with nearly ballistic behavior of Dirac electrons \cite{Heersche07,Miao07,Danneau08}. 
Even though the experimental results agree well with the ballistic theory of transport 
\cite{Katsnelson06,Tworzydlo06}, the role of metallic contacts in electronic properties 
of sub-micron graphene samples is debated \cite{Blake09}. In this paper we introduce 
a notion of metallic proximity effect in graphene and show that this effect can 
lead to a non-uniform gating of small graphene flakes. 

Despite the intense graphene studies, the scattering mechanisms, 
which limit electron mobility in the best available graphene flakes, 
are not clearly identified \cite{Ponomarenko09}. 
Still, the experimental evidence that strong impurities, 
such as adatoms and vacancies, dominate the scattering 
at low temperatures is currently building up \cite{Schedin07,Chen08,Zhou08,Elias09}. 
We demonstrate below that even the point-like impurities may introduce 
large spacial variations of electronic density at distances of the order 
of the system size from the impurity provided the quasiparticle energies are close 
to the Dirac degeneracy point. 

An impurity can be characterized by at least two length scales:
its physical size $a$ and the characteristic scattering length $\ell_s$. 
The latter is determined by the scattering cross-section and, therefore, 
depends on the energy of Dirac quasiparticles that scatter on the impurity.
A single weak impurity, i.e. the impurity with $\ell_s\ll a$, has 
practically no effect on transport and electronic density of states in ballistic sample. 
The effect becomes essential only for a substantial concentration
of weak impurities \cite{Ostrovsky06,Ostrovsky07,Schuessler09}.
In contrast, a single strong impurity, i.e. the one with $\ell_s\gg a$, can have 
a noticeable effect on transport \cite{authors09}. 

The strong impurities are of a particular importance at zero doping,
i.e. in the situation when the chemical potential in graphene is tuned 
precisely to the Dirac point. It is theoretically predicted 
that each strong impurity can form a resonant ``mid-gap'' state \cite{Wehling07},
which contributes a universal value to the conductance 
and shot noise \cite{authors09} provided the wave-length 
of the Dirac quasiparticle, $\lambda$, exceeds the distance between 
the metallic leads, $L$. The resonant conditions are equivalent 
to the limit $\ell_s\gg L$. Evidently, in this regime, the properties of graphene flakes  
are strongly influenced by both the impurities and the sample boundaries. 

The values of conductance and shot noise observed in experiments 
with sub-micron graphene flakes \cite{Miao07,Danneau08} agree with the predictions of 
ballistic theory \cite{Katsnelson06,Tworzydlo06} for $\lambda \gg L$, 
which makes us to believe that this limit is experimentally achievable. 
Both the ballistic theory of ref.~\cite{Tworzydlo06} and experiment \cite{Miao07,Danneau08} 
deal with rectangular graphene samples with large aspect ratio, $W/L \gg 1$, 
where $W$ is the sample width and $L$ is the distance between the metallic leads.
 
\begin{figure}
\centerline{\includegraphics[width=0.9\columnwidth]{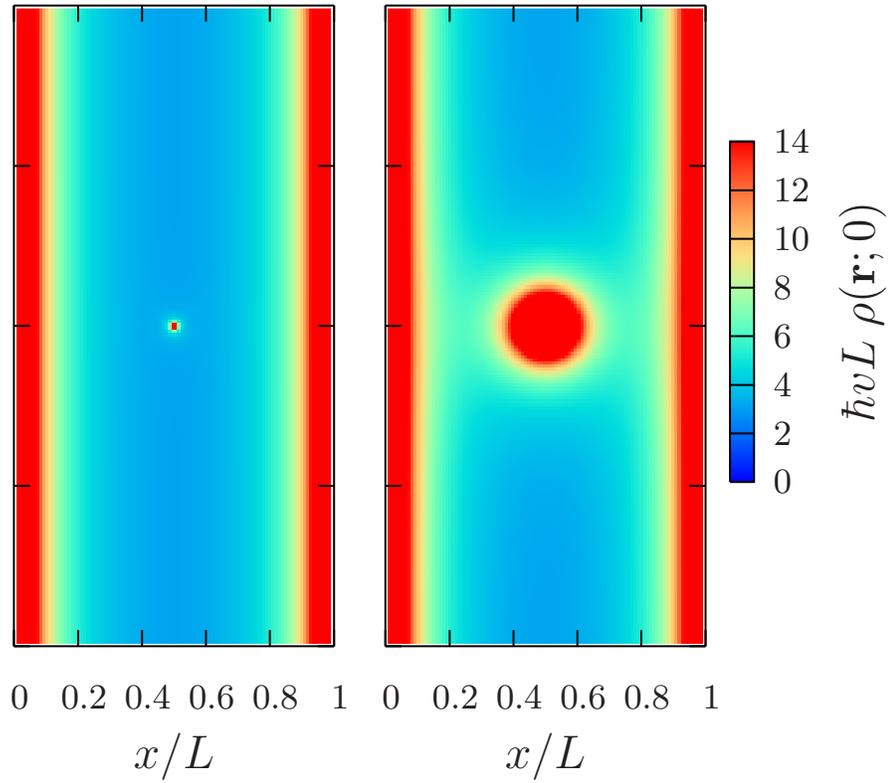}}
\caption{The density plot for the LDOS at $\ep=0$ calculated from 
the equation (\ref{final}) or out-of-resonance (left) and resonance (right) conditions. 
The physical size of impurity is chosen as $a=0.05 L$, while the effect on the LDOS
is determined by the effective length $\ell_s$. At resonance conditions, $\ell_s\gg L$,
the LDOS is strongly enhanced in the large area, which is of the order of the system size $L$.
}
\label{fig:impurity}
\end{figure}

For $\lambda \gg L$ the proximity to metal gives rise to the so-called 
pseudo-diffusive charge transport \cite{BeenakkerRMP08} in 
the model with abrupt boundary conditions at graphene-metal boundaries.
We shall demonstrate below that in this model the local density of states (LDOS) 
takes on the most simple form, $\rho_0(x)= 2[\pi \hbar v L\sin(\pi x/L)]^{-1}$, 
which corresponds to the lowest curve in figure~\ref{fig:clean}. 
The enhancement of the density near the graphene-metal interfaces 
illustrates the metallic proximity effect. The LDOS in the presence of a single impurity 
is shown in figure~\ref{fig:impurity} with a contour plot for resonance 
and off-resonance conditions. Such a strong spacial 
dependence of the LDOS suggests that, in the presence of electron-electron
interactions, the sample is not uniformly charged by the gate electrode.

\section{Clean sample}

The LDOS is defined by the general formula, 
$\rho(\bb{r};\ep) = -\pi^{-1}\im \tr G^R(\bb{r},\bb{r};\ep)$,
where the retarded Green's function yields the matrix equation
\begin{equation}
\label{green}
(\ep+i0-\mu(x)-H) G^R(\bb{r},\bb{r}')=\delta(\bb{r}-\bb{r}'),
\end{equation}
with $H=H_0+U(\bb{r})$. In this paper the Dirac Hamiltonian of ideal graphene, 
$H_0=-i\hbar v \bb{\sigma}\bb{\nabla}$, is taken in the valley symmetric representation,
where $v=10^6$\,m/s is an effective "speed of light". The impurity potential, $U(\bb{r})$,
is considered to be a scalar in the Dirac equation, which assumes that the corresponding 
physical potential is smooth on the scale of the graphene lattice. The vector 
$\bb{\sigma}=(\sigma_x,\sigma_y)$ is a two-component vector of Pauli matrices 
and the energy, $\ep$, determines the wave-length of free Dirac quasiparticle 
as $|\ep|=h v/\lambda$. 

Following earlier works we describe the metal-graphene interfaces at $x=0, L$  
with a step function model  
\begin{equation}
\label{leads}
\mu(x)=\lt\{
\ba{cc} - hv/\lambda_F,&\quad x<0,\; x>L \\
0, & \quad 0<x<L \ea\rt. ,
\end{equation}
where $\lambda_F$ is regarded as the Fermi wave length in the metallic lead. 
We naturally assume that $\lambda_F\ll \min\{ L, \lambda\}$.  
In order to avoid a detailed discussion of the sample edges 
we impose periodic boundary conditions in $y$ direction. 

We start with a solution of equation~(\ref{green}) in the absence of impurity potential. 
This solution is most easily obtained in the channel representation, which is defined by 
the Fourier transform, 
\begin{equation}
G^R(\bb{r},\bb{r}')=\frac{1}{W}\s_q e^{iq(y-y')} G_{q}^R(x,x'),
\end{equation}
where the summation runs over the discrete values 
of transversal momentum, $q_n=2\pi n/W$. Selecting the solutions which decay 
in the leads we obtain the effective boundary conditions on graphene-metal interfaces 
in the limit $\lambda_F\ll \min\{ L, \lambda\}$ \cite{authors09},
\begin{eqnarray}
\nonumber
(1,1)G_{q}^R(0,x') = 0, &\qquad& (1,-1)G_{q}^R(L,x')=0,\\
G_{q}^R(x,0)(1,-1)^T=0, &\qquad & G_{q}^R(x,L)(1,1)^T=0.
\label{boundary}
\end{eqnarray}
Using these boundary conditions the Green's function inside the clean sample 
is readily found as
\begin{equation}
\label{GR0}
G^{R}_{0,q}(x,x';\ep)=\frac{-i \lt(F_1(u_1)+F_2(u_2)\rt)}{2\hbar v \,\cos \theta\, \cos\Phi},
\end{equation}
where $u_1=\Phi-k|x-x'|$, $u_2=k(L-x-x')$ and 
\begin{eqnarray}
\label{defs}
&& k=\sqrt{(\ep-i0)^2-(\hbar v q)^2},\quad e^{i\theta}=\hbar v (k+iq)/\ep,\\ 
&& e^{i\Phi}=e^{ikL}(1-e^{i\theta})/(1+e^{i\theta}).
\end{eqnarray}
In equation~(\ref{GR0}) the following definitions were used,
\begin{eqnarray}
F_1(u) &= &
\lt(\ba{cc} 
-i\sin u & \eta \cos(u+\eta\, \theta) \\
\eta \cos(u-\eta\, \theta) & -i\sin u
\ea\rt),\\
F_2(u) &= &
\lt(\ba{cc} 
-i\sin(u+\theta ) & -\cos u \\
\cos{u} & \phantom{-}i\sin(u-\theta)
\ea\rt),
\end{eqnarray}
where $\eta=\mathrm{sign}\,(x-x')$. The quantity, $k$, introduced 
in equation~(\ref{defs}) has a meaning of $x$ component 
of the quasiparticle momentum, which can take on complex values. 

\begin{figure}
\centerline{\includegraphics[width=0.9\columnwidth]{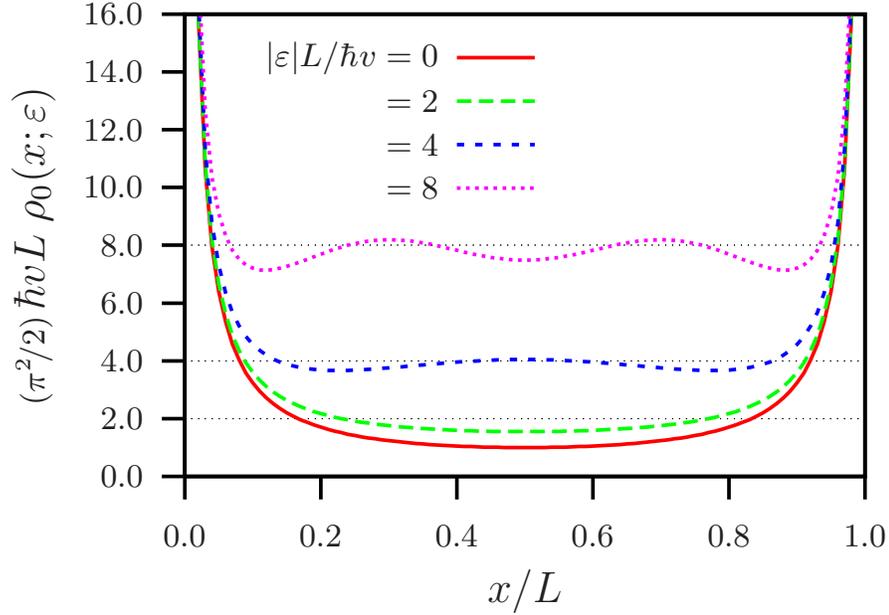}}
\caption{The local density of states of Dirac quasiparticles 
from the equation (\ref{dens_clean}) in ballistic graphene sample 
with metal boundaries and $W/L=10$. Different curves correspond 
to different energy values $\ep L/\hbar v= 0,2,4,8$. In the middle of the sample
(i.e. for $\lambda \ll x \ll L-\lambda$, where $\lambda=2 \pi \hbar v/|\ep|$)
the LDOS is approximately given by $\rho_0 = |\ep|/(\pi \hbar v)^2 $ (dotted lines). 
Near the metal boundaries (with  $\lambda_F=0$) the LDOS behaves 
as $\rho_0=[\pi^2\hbar v d]^{-1}$, where $d=\min\{x,L-x\}$.
}
\label{fig:clean}
\end{figure}

Thus, the LDOS of a clean system is given by the expression
\begin{equation}
\label{dens_clean}
\rho_0(x;\ep)=4\,\im 
\s_q \frac{\sin\theta\cos(k(L-2x))+\sin\Phi}{\pi \hbar v W\,\cos\theta \,\cos\Phi},
\end{equation}
where the factor $4$ takes into account the spin and valley degeneracies. 
The LDOS of the clean setup has no dependence on $y$ due to the translational symmetry 
of the system in this direction. The $x$-dependence of the LDOS  
is plotted in figure~\ref{fig:clean} for $W/L=10$ and several values of energy. 

The expression (\ref{dens_clean}) can be approximated in the doped graphene, $\lambda \ll L$, 
by averaging over the Friedel oscillations on the scale $\lambda$. Away from the boundaries,
$d\equiv\min\{x, L-x\} \gg \lambda$, the evanescent modes, which are characterized by complex 
momenta $k$, can be disregarded and the averaging over the Friedel oscillations can be replaced 
by the averaging over the phase $\phi=kL$. For $W\gg L$, the averaged LDOS value 
is given by $\bar{\rho}_0  = 2|\ep|/(\pi\hbar v)^2$, which is shown in figure~\ref{fig:clean} 
with dotted lines. Near the sample edge, $d\equiv \min\{x,L-x\} \ll \lambda$, 
the density is dominated by the metallic proximity effect, so that it diverges at
the metal-graphene interfaces in the limit $\lambda_F\to 0$.
For $d\ll \lambda=hv/|\ep|$ and $W\gg L$, the LDOS does not depend on energy 
and can be approximated as $\rho_0 =2[\pi^2\hbar v d]^{-1}$. 

In contrast, for $\lambda \gg L$, the LDOS is fully determined by the evanescent modes. 
In particular, for $\ep=0$ ($\lambda \to \infty$), the result can be obtained 
by taking $k=-iq$, $\theta=+i \infty$, and $\Phi=-iqL$ in equation~(\ref{dens_clean}),
which leads to the expression
\begin{equation}
\label{rho0general}
\rho_0(x;0)=\frac{4}{\pi\hbar v W} \s_q \frac{\cosh \lt(q(L-2x)\rt)}{\cosh(qL)},
\end{equation}
In the limit $W\gg L$, the sum over the transversal momenta, $q$, can be replaced 
by the integration, with the result
\begin{equation}
\label{refLDOS}
\rho_0(x)=\frac{2}{\pi \hbar v(L\sin(\pi x/L)+\lambda_F/4)},
\end{equation}
where the small term $\lambda_F/4$ in the denominator is effective only at the edges,
$x=0$ and $x=L$, in which case the sum (\ref{rho0general}) equals the number 
of channels in the lead.

The result (\ref{refLDOS}) shows that the LDOS at $\ep=0$ is dominated, 
in the limit $W\gg L$, by the states originating in the leads. 
These boundary states, which are also referred to as the evanescent modes, are responsible for 
the so-called pseudo-diffusive transport regime in this setup \cite{BeenakkerRMP08}, 
which is characterized by universal full counting statistics \cite{authors09}. 
We see, therefore, that the strong metallic proximity effect is the natural property 
of the abrupt interface model (\ref{leads}), which is compatible 
with the transport measurements of ref.~\cite{Danneau08}. 

\section{Effect of impurity}

In this Section we consider the effect of a scalar point-like impurity 
on the LDOS at $\ep=0$ for a rectangular graphene sample with $W\gg L$.
The calculation of the Green's function is analogous to the one presented 
in ref.~\cite{authors09} for the charge transport and can be carried 
out straightforwardly.

We shall start with the real space representation 
of the free Green's function (\ref{GR0}) at $\ep=0$, 
that reduces, in the limit $W\gg L$, to \cite{authors09}  
\begin{equation}
\label{real}
G^{R}_{0,q}(\bb{r},\bb{r}')=\frac{-i}{4\hbar v}
\lt(\ba{cc} C_+(\bb{r},\bb{r}') & C_-(\bb{r},\bb{r}') \\
C_-^{*}(\bb{r},\bb{r}') &  C_+^{*}(\bb{r},\bb{r}')
\ea\rt),
\end{equation}
where 
\begin{equation}
\label{C}
C_{\pm}(\bb{r},\bb{r}')=\lt[L \sin \lt(\pi (x\pm x'+i(y-y')) /2L \rt) + \lambda_F/4 \rt]^{-1}.
\end{equation}
One can see from the equations~(\ref{real},\ref{C}) 
that the free Green's function obeys the chiral symmetry
\begin{equation}
\lt[G^R_0(\bb{r'},\bb{r})\rt]\h =-\sigma_z G^R_0(\bb{r},\bb{r}')\sigma_z=G^{A}_{0}(\bb{r},\bb{r}'),
\end{equation}
which is, therefore, preserved in the abrupt interface model 
described by equation~(\ref{leads}).

One can also see from equation (\ref{real}) that the Green's function 
at the coinciding arguments can be decomposed into the sum 
of the free Green's function in an infinite system, $g$, 
and the proximity-induced part, $G_\mathrm{reg}$, such that
\begin{equation}
\label{decompose}
\lim_{\bb{r}'\to\bb{r}}G^R_0(\bb{r},\bb{r}') =  g(\bb{r},\bb{r}')+G_\mathrm{reg}(\bb{r}),
\end{equation}
where $g$ is off-diagonal and singular,
\begin{equation}
g(\bb{r},\bb{r}')=-\frac{i}{2\pi \hbar v}\frac{\bb{\sigma}(\bb{r}-\bb{r'})}{|\bb{r}-\bb{r}'|^2},
\end{equation}
while $G^R_{\mathrm{reg}}(\bb{r})$ is proportional to the product 
of the LDOS (\ref{refLDOS}) and the unit matrix,
\begin{equation}
G^R_{\mathrm{reg}}(\bb{r})=(-i\pi/4)\rho_0(x).
\end{equation}

In order to simplify the calculation we choose the impurity potential in the form of a disk
with sharp boundaries, $U(\bb{r})=u\; \theta(a-|\bb{r}-\bb{r}_0|)$, where 
the theta function, $\theta(x)$, equals $1$ for $x\geq 0$ and zero otherwise.
We also take the limit of point-like impurity, $a/L \ll 1$. In this limit 
the full T-matrix of the impurity can be written, in the symbolic notations, as
\begin{equation}
\bb{\mathrm{T}}(\bb{r}_0)=U\frac{1}{1-G_0^R U}=T\frac{1}{1+ (i \pi/4)\rho_0(x_0) T},
\end{equation}
where we used the decomposition (\ref{decompose}) and introduced the 
reduced T-matrix, $T= U [1-g U]^{-1}$, which is represented, in the detailed notations, 
by the infinite sum
\begin{equation}
T=\s_{s=1}^{\infty} \int \p_{p=1}^s d^2\bb{r}_p\; 
U(\bb{r}_1)g(\bb{r}_1,\bb{r}_2)
U(\bb{r}_2) \dots g(\bb{r}_{s-1},\bb{r}_s)U(\bb{r}_s).
\label{Tinf}
\end{equation}

The reduced T-matrix given by equation (\ref{Tinf}) does not contain 
any information on the metallic leads and sample boundaries. 
Therefore, it can be obtained from the consideration
of scattering of free Dirac quasiparticles on a single disk impurity. 
This problem has been addressed by several authors \cite{Hentschel07,Basko08,authors09}.
The result in the s-wave approximation and in the limit $\ep\to 0$ takes the form
\begin{equation}
\label{Tresult}
T=2\pi \hbar v \ell_s,\qquad \ell_s= a \frac{J_1(ua/\hbar v)}{J_0(ua/\hbar v)},
\end{equation}
where $J_n$ is the Bessel function and the T-matrix, $T$, is
proportional to the unit matrix in spin, valley, and sublattice space.

The equation (\ref{Tresult}) completes the calculation of the Green's function in 
the presence of a single impurity, 
\begin{equation}
\label{Gfinal}
G^R(\bb{r},\bb{r'})=G^{R}_0(\bb{r},\bb{r'})+G^{R}_0(\bb{r},\bb{r}_0)
\bb{\mathrm{T}}(\bb{r}_0) G^{R}_0(\bb{r}_0,\bb{r}),
\end{equation}
where $\bb{r}_0=(x_0,y_0)$ specifies the impurity location. 
We stress that the characteristic impurity size, 
$\ell_s$, determined by the reduced T-matrix, strongly depends on the impurity 
potential, $u$, and diverges at the resonance. 

From equation (\ref{Gfinal}) we find the LDOS at the point $\bb{r}$, 
which can be written as
\begin{equation}
\label{final}
\rho(\bb{r})=\rho_0(x)+ \rho_0(x_0)
\frac{\lt| C_-(\bb{r},\bb{r}_0)\rt|^2-\lt|C_+(\bb{r},\bb{r}_0)\rt|^2}{(2/\pi\ell_s)^{2} 
+\lt[\pi\hbar v \rho_0(x_0)/2\rt]^2}.
\end{equation}
This result is plotted in the figure~\ref{fig:impurity} for an impurity placed 
in the middle of the sample, $x_0=L/2$. 

For the resonant conditions, $J_0(ua/\hbar v)=0$, (i.e. for $\ell_s\to \infty$) 
the disk impurity supports a quasi-bound state at $\ep=0$ \cite{Bardarson09,authors09}
with only one pseudo-spin component being localized. 
In this case, the LDOS is strongly enhanced at distances of the order 
of $L$ from the impurity. Away from the resonant condition, 
the effect of a single impurity on the LDOS is predictably small. 
It was argued on the basis of the DFT analysis \cite{Wehling07} 
that some single-atom impurities such as hydrogen naturally 
create the mid-gap quasi-bound states.
The resonance condition for such scatterers are always fulfilled 
near the Dirac point. Therefore, even few impurities 
in the sample will have a great impact on the LDOS. We expect that this effect 
can be observed with the STM measurements. 

Finally we note that the denominator in the second term of equation~(\ref{final}) 
coincides with the denominator in the impurity contribution to the conductance 
in the same setup \cite{authors09}. Near the resonance, the second term 
in equation~(\ref{final}) has a Lorentzian dependence on the position 
of the impurity level with respect to the Dirac point. 
This characteristic dependence reflects the resonant nature of charge transport that originates 
in the impurity assisted tunneling in undoped graphene \cite{Titov07}. 

\section{Discussion}

We found from equation~(\ref{dens_clean}) that the LDOS in ballistic rectangular 
sample with $W\gg L$ can be approximated as
\begin{equation}
\label{rough}
\rho_0(x;\ep)=\frac{2}{(\pi\hbar v)^2}\max\lt\{|\ep|,\frac{\hbar v}{d}\rt\},
\end{equation}
where $d=\min\{x,L-x\}$ is the distance to the closest graphene-metal interface.
This result shows that the metal proximity effect is fully dominating 
the LDOS for energies close to the Dirac point, $|\ep|< \hbar v/L$. 
It also demonstrates that the LDOS is never vanishing in graphene with metal boundaries.

From equation~(\ref{rough}) we readily estimate the integrated density of states as 
\begin{equation}
\label{integrated}
\nu_0(\ep)=\frac{2 W}{\pi^2\hbar v}\lt(\pi L/\xi+ 2\ln(\xi/\lambda_F)\rt),
\end{equation}
where $\xi=\min\{L,\pi\hbar v/|\ep|\}$. Thus, for $|\ep|< \hbar v/L$, 
the integrated density increases logarithmically with the system size, $L$, due to 
the electronic states at the metal-graphene interfaces. Peculiar transport properties 
of ballistic graphene \cite{BeenakkerRMP08,authors09} are originating 
in the metallic proximity effect.

Similarly, the presence of the resonant impurity has a great effect on the LDOS 
near the Dirac point since the corresponding electronic states localize not only on the metal interfaces 
but also on the impurity. This effect leads to a highly inhomogeneous electronic density profile
for $|\ep|< \hbar v/L$, which is illustrated in figure~\ref{fig:impurity}. 

This result implies that electron-electron interactions may prevent 
the uniform gating of the sample in the vicinity of the Dirac point. 
Assuming that the interaction potential is given by $V(\bb{r})$ we can 
estimate, up to a constant value, an additional potential profile 
in the single-electron problem (\ref{green}) as 
\begin{equation}
\label{int}
\phi(\bb{r})=\il_0^L\!\! dx' \!\! \il_0^W \!\! dy'\, 
V(\bb{r}-\bb{r}')\!\! \il_{-E_c}^{E_c}\!\! d\ep\, 
\rho(x';\ep)f(\ep), 
\end{equation}
where $f(\ep)$ is the Fermi distribution function. Since only the spatially 
varying part of the potential $\phi(\bb{r})$ is relevant, we can let $E_c=\hbar v/L$.
Finally the LDOS and the potential profile, $\phi(\bb{r})$, have to be found 
self-consistently from equation (\ref{green}) with $\mu(x)\to \mu(x)+\phi(\bb{r})$
and equation (\ref{int}). The detailed analysis of the non-uniform gating is, however, 
beyond the scope of the present paper.

\section*{Acknowledgements}

The work was supported by the DFG -- Center for
Functional Nanostructures and by EUROHORCS/ESF EURYI Award (I.V.G.).
The discussions with Alexander Mirlin, Pablo San-Jose, Henning Schomerus, 
and Alexander Schuessler are gratefully acknowledged.

%
%


%
%


\end{document}